# An Algorithmic Perspective on Some Network Design, Construction and Analysis Problems

**Mugurel Ionut Andreica, Politehnica University of Bucharest**
**Mihai Aristotel Ungureanu, Romanian-American University**
**Romulus Andreica, Commercial Academy of Satu Mare**
**Angela Andreica, Commercial Academy of Satu Mare**

**Abstract:** *Efficient network design, construction and analysis are important topics, considering the highly dynamic environment in which data communication occurs nowadays. In this paper we address several problems concerning these topics from an algorithmic point of view.*

## 1. Introduction

Efficient network design and analysis techniques are strongly required in the highly dynamic environment in which data transfers occur nowadays. In this paper we focus on several algorithmic aspects of network design and analysis. We consider several constrained network design and (re)construction problems (Section 2), and several edge and vertex classification problems (Section 3). In Section 4 we discuss related work and we conclude.

## 2. Optimal Network Design and (Re)construction Problems

We consider $N \geq 3$ points in the plane, whose coordinates $(x(i),y(i))$ are not given. Every point $i$ has a weight $w(i)>0$ ($1 \leq i \leq N$). The arrangement of points is considered *balanced* if for every $3$ distinct points $A$, $B$ and $C$, we have $w(A)+w(B)+w(C)=AreaTri(A,B,C)$ (*AreaTri(A,B,C)* is the area of the triangle *ABC*). We want to compute the coordinates of the points, such that the resulting arrangement is balanced. Moreover, we have $w(N-1)=w(N)$. We first notice that there is no solution for $N \geq 5$. Thus, we only need to handle the cases $N=3$ and $N=4$. For $N=3$ we can choose: $x(2)=y(2)=0$, $x(3)=2$, $y(3)=0$, $x(1)=0$, $y(1)=w(1)+w(2)+w(3)$. The triangle has a right angle at $(0,0)$. The two perpendicular sides have lengths

$C_1=w(1)+w(2)+w(3)$ and $C_2=2$. Thus, the triangle's area is $(C_1 \cdot C_2)/2=w(1)+w(2)+w(3)$. For $N=4$ we have 2 subcases. In the first subcase we have $w(1)=w(2)$ (apart from $w(3)=w(4)$). In this subcase we will choose the coordinates of the points *3* and *4* at *(0,0)* and *(2,0)*. The coordinates of point *1* will be *(0, w(1)+w(3)+w(4))*, and those of point 2 will be *(2·(w(1)+w(2)+w(4))/(w(1)+w(3)+w(4)), w(1)+w(3)+w(4))*. The *4* points form a trapezoid whose parallel sides are the segments connecting the points *1-2* and *3-4*. Triangles *1-3-4* and *2-3-4* have heights of *(w(1)+w(3)+w(4))*, perpendicular on sides of length *2*. Triangles *1-2-3* and *1-2-4* have heights equal to *(w(1)+w(3)+w(4))* and perpendicular on a side with length *2·(w(1)+w(2)+w(4))/(w(1)+w(3)+w(4))*; thus, their area is *w(1)+w(2)+w(4)=w(1)+w(2)+w(3)*. The second subcase for *N=4* occurs when *w(1)+w(2)=2·w(4)*. In this subcase, the coordinates of the points *3* and *4* will be *(0,0)* and *(2,0)*, and those of the points *1* and *2* will be *(1, (w(1)+w(3)+w(4)))* and *(1, -(w(2)+w(3)+w(4)))*. We notice that the *2* subcases for *N=4* correspond to the situations in which the points *1* and *2* are on the same side or on opposite sides of the segment *3-4*. We notice that for $N \geq 5$ (or *N=4* and weights which do not satisfy the conditions of the two subcases) we cannot have a solution, because: if we have *4* points, then either the segment *1-2* is parallel to the segment *3-4*, or the segment *1-2* intersects the segment *3-4* at its midpoint. For $N \geq 5$, one of these *2* situations should occur for the pairs *(1,2)*, *(1,3)* and *(2,3)* (relative to the segment *4-5*, respectively to the midpoint of the segment *4-5*). This indicates that either the area of the triangle *1-2-3* is *0* (but *w(1)+w(2)+w(3)>0*), or that one of the points *1*, *2* or *3* (let this point be *P*) is the midpoint of the segment *4-5*, in which case the area of the triangle *4-5-P* would be *0* (but *w(4)+w(5)+w(P)>0*).

The 2[nd] problem considers the reconstruction of a polygon with a known number *N* of vertices, whose edges may self-intersect. Let's consider that vertex *i* of the polygon has coordinates *(x(i), y(i))* *(1≤i≤N)*. On the support line of every edge *(i, i+1)* of the polygon (where *i+1=1*, for *i=N*), a point *p(i)* is chosen at distance *t(i)·D(i,i+1)* away from vertex *i* and towards vertex *i+1*. *D(i,i+1)* is the length of the edge *(i,i+1)* and *t(i)≠0*. The coordinates of the point

$p(i)$ are $(xp(i), yp(i))$. We want to reconstruct the polygon from the $N$ points $p(i)$. The coordinate $xp(i)$ ($yp(i)$) of a point $p(i)$ can be written as follows, based on the coordinates of the vertices $i$ and $i+1$: $xp(i)=x(i)+t(i)\cdot(x(i+1)-x(i))=(1-t(i))\cdot x(i)+t(i)\cdot x(i+1)$  ($yp(i)=(1-t(i))\cdot y(i)+t(i)\cdot y(i+1)$). Based on these equations we will write the coordinate $x$ ($y$) of every vertex as a linear function of the coordinate $x(1)$ ($y(1)$). We have $x(i)=ax(i)\cdot x(1)+bx(i)$ ($y(i)=ay(i)\cdot y(1)+by(i)$). For $i=1$ we have $ax(1)=1$, $bx(1)=0$, $ay(1)=1$, and $by(1)=0$. For $2\leq i\leq N+1$, we will obtain $x(i)$ ($y(i)$) from the equation of point $p(i-1)$: $x(i)=(xp(i-1)-(1-t(i-1))\cdot x(i-1))/t(i-1)=(t(i-1)-1)/t(i-1)\cdot x(i-1)+xp(i-1)/t(i-1)$ (the equation for $y(i)$ is similar; we just replace $xp(i-1)$ by $yp(i-1)$, and $x(i-1)$ by $y(i-1)$). From here we obtain: $ax(i)=(t(i-1)-1)/t(i-1)\cdot ax(i-1)$, $bx(i)=(t(i-1)-1)/t(i-1)\cdot bx(i-1)+xp(i-1)/t(i-1)$ (the equation for $y(i)$ is similar – we just replace the character "$x$" by "$y$" in the previous equation). After computing the values for the vertex $N+1$ (which is, in fact, vertex $1$), we will write the equations: $ax(N+1)\cdot x(1)+bx(N+1)=x(1)$ => $x(1)\cdot(ax(N+1)-1)=-bx(N+1)$ (similarly, $y(1)\cdot(ay(N+1)-1)=-by(N+1)$). We will determine the coordinates $x$ and $y$ of the polygon vertices independently. We will present the method only for the coordinate $x$, because it is identical for the coordinate $y$ (we just replace $x$ by $y$). If $(ax(N+1)-1)\neq 0$, then we obtain $x(1)=-bx(N+1)/(ax(N+1)-1)$, and then we will compute the coordinates of the other vertices: $x(i)=ax(i)\cdot x(1)+bx(i)$. If $(ax(N+1)-1)=0$, then we have 2 subcases. In the first subcase, $bx(N+1)=0$; thus, we can choose any value for $x(1)$, afterwards obtaining the $x$ coordinates of the other vertices. In the second subcase, $bx(N+1)\neq 0$ and the problem admits no solution. If we allow $t(q)=0$ (i.e. $xp(q)=x(q)$ and $yp(q)=y(q)$), then we split the points $q$ into maximal chains $[i..j]$, such that $t(j)=t(i-1)=0$ and $t(i)$, $t(i+1), \ldots, t(j-1)$ are non-zero (with $q+1=1$, for $q=N$, and $q-1=N$, for $q=1$). For every chain $[i..j]$, we write $x(k)$ ($y(k)$) as a linear function $fx(k)$ ($fy(k)$) of $x(i)$ ($y(i)$) ($k$ belongs to $[i..j]$), like we did before, when we considered linear functions of $x(1)$ ($y(1)$). Since $x(j)=xp(j)$ ($y(j)=yp(j)$), we compute $x(i)$ ($y(i)$) from $fx(j)=xp(j)$ ($fy(j)=yp(j)$) and, based on them, all the other values $x(k)$ and $y(k)$ of the points $k$ of the chain $[i..j]$. The time complexity of the algorithm is $O(N)$ in any case.

In the *3rd* problem we consider a triangle *ABC*, in which the length of the side *AB* (denoted by *lc*), the length of the side *AC* (denoted by *lb*) and the length of the median *AM* (denoted by *lm*) are given. *AM* is the segment which connects the vertex *A* of the triangle *ABC* to the midpoint of the segment *BC*. We want to compute the coordinates of the *3* vertices. We will binary search the length of the segment *BC*. Let's assume that we considered this length to be *2·a*. We will test if this length is valid. We will consider that the coordinates of the vertex *B* are *(0,0)*, and those of vertex *C* are *(2·a, 0)*. This way, the midpoint of the segment *BC* is at coordinates *(a,0)*. If *lm+a<lc* or *lm+a<lb*, then the length *2·a* is too small. Otherwise, we will binary search, between *0* and $\pi$, the angle *AMB* (determined by the points *A*, *M* and *B*). Let's assume that, during the binary search, we selected an angle *alpha*. We compute the corresponding length of the side *AB*, as being *c'=sqrt(lm²+a²-2·lm·a·cos(alpha))* (we denote by *sqrt(x)* the square root of *x*). If *c'<lc*, then *alpha* is too small and we will consider a larger angle next; otherwise, we will consider a smaller angle next. The binary search of the angle will end when the length of the search interval becomes smaller than a constant *ueps*. After finding the angle *alpha* for which the length of the side *AB* is equal to *lc*, we will compute the length of the side *AC* corresponding to this angle: *b'=sqrt(lm²+a²-2·lm·a·cos($\pi$-alpha))*. If *b'<lb*, then the length *2·a* chosen within the binary search for the length of the side *BC* is too small and we will search for a larger value; otherwise, we will search for a smaller value. We finish the binary search when the length of the search interval becomes smaller than a predefined constant *leps*. The algorithm contains *2* nested binary searches and its complexity is *O(log(LMAX)·log(UMAX))*; *LMAX* is the maximum possible length of *BC* and *UMAX=$\pi$*. Alternatively, we can solve the system of equations (where we used *cos($\pi$-alpha)=-cos(alpha)*): *(1) lc²=lm²+a²-2·lm·a·cos(alpha)* ; *(2) lb²=lm²+a²+2·lm·a·cos(alpha)*. By adding *(1)* and *(2)* we obtain *a=sqrt((lc²+lb²)/2-lm²)*. After replacing the value of *a* in *(1)*, we can also compute *cos(alpha)*.

## 3. Classifying Edges and Vertices Relative to Matchings

We consider a bipartite multigraph with *n* vertices overall on its left and right parts. Every edge *(u,v)* (*u* on the left side and *v* on the right side) has a cost $c(u,v) \geq 0$. We want to classify every edge (vertex) of the multigraph in one of the following *3* categories: *1)* it belongs to every minimum cost maximum matching ; *2)* it belongs to at least one minimum cost maximum matching (but not to all of them) ; *3)* it belongs to no minimum cost maximum matching. We start by computing a minimum cost maximum matching *M*, in $O(n^3)$ time. Obviously, the edges (vertices) from category *1* must be among the edges (endpoints of the edges) of *M*. Then, we will construct a directed multigraph $G_1$, as follows. We direct every edge *(u,v)* which belongs to *M* from *v* to *u* and we assign it a cost $c'(v->u)=-c(u,v)$. We direct every edge *(u,v)* which does not belong to *M* from *u* to *v* and we assign it a cost $c'(u->v)=c(u,v)$. Then, we add an extra vertex *S* and all the zero-cost directed edges *(S->u)* and *(v->S)* (*u* and *v* on the left side), where *u* is a vertex such that no edges adjacent to it belong to *M*, and *v* is a vertex such that there is an edge adjacent to it in *M*. After constructing this multigraph, we will compute the set $E_1$ of edges which belong to at least one zero-cost cycle (considering the costs *c'*), i.e. a cycle whose sum of edge costs is *0*. This test can be performed in $O(n^3)$ time using a standard method [1] for multigraphs with no negative cost cycles (as is our case), i.e. without cycles whose sum of edge costs is negative: we make all the edge costs non-negative and we consider the directed multigraph *G'* with the zero cost edges only; the edges of *G'* which connect two vertices from the same strongly connected component of *G'* belong to a zero-cost cycle in $G_1$. Then, we construct another directed multigraph $G_2$, starting from the initial multigraph. We direct the edges and assign them costs just like before. This time we will add a vertex *T* instead of *S*, together with all the zero-cost directed edges *(v->T)* and *(T->u)* (*u* and *v* on the right side), where *v* is a vertex which has no edge adjacent to it belonging to *M*, and *u* is a vertex which has one edge adjacent to it belonging to *M*. We then compute the set of edges $E_2$ which belong to at least one zero-cost cycle in $G_2$. Let $E_{1,a}$ be the set of edges *(S->v)* from $E_1$ and $E_{1,b}$ the set of edges *(v->S)* from $E_1$, and let $E_{2,a}$ be the set of edges *(v->T)* from $E_2$ and $E_{2,b}$ the set of edges *(T->v)* from $E_2$. After this, we set $E_1=(E_1 \backslash E_{1,a}) \backslash E_{1,b}$ and $E_2=(E_2 \backslash E_{2,a}) \backslash E_{2,b}$ and we interpret the edges in $E_1$ and $E_2$ as being undirected. The edges from category *1* are those from the set $M \setminus (E_1 \cup E_2)$; those from category *2* form the set $(E_1 \cup E_2)$. The edges from category *3* are the other edges of the multigraph (which do not belong to $M \cup (E_1 \cup E_2)$ ). Let *Q(S)* be the vertex set containing all the endpoints of the edges in the set of edges *S*. The set of vertices from category *1* is *(Q(M)\Q(E_{1,b}))\Q(E_{2,b})*, from category *2* is $(Q(E_{1,a}) \cup Q(E_{1,b}) \cup Q(E_{2,a}) \cup Q(E_{2,b})) \setminus \{S,T\}$ ; the remaining vertices belong to category *3*. Thus, in $O(n^3)$ time, we can classify all the edges (vertices) of the multigraph in the *3* categories.

We will now consider a similar problem, but for bipartite multigraphs without edge costs. We want to classify the edges (vertices) of the multigraph as: *1)* belonging to every maximum matching ; *2)* belonging to at least one maximum matching (but not all of them) ; *3)* belonging to no maximum matching. As before, we start by computing a maximum matching *M*, in $O(n^{2.5})$ time. Then, we construct the multigraphs $G_1$ and $G_2$ like before, except that we do not assign costs to the edges. We compute the strongly connected components of $G_1$ and $G_2$ (in $O(n^2)$ time) and then we compute the sets of edges $E_1$ and $E_2$, consisting of the edges between two vertices *u* and *v* which belong to the same strongly connected component in $G_1$ and, respectively, $G_2$. We also compute $QV_1$ ($QV_2$) as the set of vertices *u* such that the edge *(u,S)* *((u,T))* or *(S,u)* *((T,u))* exists in $E_1$ ($E_2$). Then, we remove the edges *(S,\*)* and *(\*,S)* from $E_1$, the edges *(T,\*)* and *(\*,T)* from $E_2$, and then we interpret the edges in $E_1$ and $E_2$ as being undirected. Just like before, the set of edges in category *1* is $M \setminus (E_1 \cup E_2)$, the set of edges in category *2* is $(E_1 \cup E_2)$ and the edges in category *3* are those edges which do not belong to $M \cup (E_1 \cup E_2)$. The set of vertices from category *1* is *(Q(M)\QV_1)\ QV_2* and from category *2* is $(QV_1 \cup QV_2)$; the remaining vertices are in category *3*. The algorithm is dominated by the computation of a maximum matching, and its complexity is $O(n^{2.5})$.

A related problem (whose solution was mentioned to us by C. Negruseri) asks for classifying the edges of a connected multigraph with *N* vertices and *M* edges as follows: *1)* belonging to every minimum spanning tree (MST) ; *2)* belonging to at least one MST

(but not all) ; *3)* belonging to no MST. We will sort the edges in ascending order, according to their cost, and we will then consider groups of edges with equal costs, in increasing order of the costs. We will use the disjoint sets data structure, which provides the functions *Find(x)* and *Union(x,y)*, for maintaining the connected components of the graph's vertices. Let's assume that we reached the group of all edges with cost *C*. Let *QV(C)* be the set of endpoints of the edges in the group. We will construct a multigraph *G(C)* whose set of vertices is the set *{Find(x)|x∈ QV(C)}*; every edge *(x,y)* from the current group induces an edge *(Find(x), Find(y))* (possibly a loop, if *Find(x)=Find(y)*) in *G(C)*. *Find(x)* denotes the representative of the connected component containing node *x*, in the graph *G'(C)*=the graph containing all the *N* vertices and all the edges with cost strictly smaller than *C*. The edges *(x,y)* inducing critical edges (bridges) in *G(C)* belong to category *1*; the edges inducing non-bridge edges which connect two different nodes in *G(C)* belong to category *2*; the edges inducing loops (i.e. those edges which connect a node to itself) in *G(C)* belong to category *3*. After considering a group of edges with cost *C*, for every edge *(x,y)* in the group, we set that *x* is in the same connected component as *y* (we call *Union(x,y)*). Computing the critical edges (bridges) of a multigraph with $n_C$ vertices and $m_C$ edges can be done in $O(n_C+m_C)$ time (since we consider only those vertices which are adjacent to at least one of the $m_C$ edges => $n_C=O(m_C)$). The different stages of the algorithm have the following complexities: $O(M \cdot log(M))$ for sorting the edges (or $O(M+CMAX)$ if the edge costs are integer and are bounded by a small integer value *CMAX*); $O(M \cdot log_*(N))$ for all the *Find(\*)* and *Union(\*,\*)* calls; $O(M)$ overall for computing the critical edges (bridges) for all the groups of equal-cost edges. If we drop the costs of the edges, then every bridge (critical edge) of the multigraph belongs to every spanning tree and every other edge belongs to at least one spanning tree (i.e. for every edge *(i,j)* there is a spanning tree *ST* which contains it).

We will now consider a directed flow network with designated *source* and *sink* vertices (and possibly parallel edges) and we want to classify the network edges as being *upward critical* or not (an edge *(u->v)* is upward critical if by increasing its capacity by *1* unit and leaving the capacities of the other edges the same, the maximum flow in the network increases). First, we compute a maximum flow from the source(s) to the sink(s). Thus, every directed edge *(u->v)* has a flow *f(u->v)* on it, such that $f(u\text{->}v) \leq cap(u\text{->}v)$ (where *cap(u->v)* is the capacity of the edge). We perform a breadth-first search (*BFS*) starting from the source(s) and considering only those edges *(u->v)* with *f(u->v)<cap(u->v)*. We mark the reachable vertices *i* by $reachable_1(i)=true$ (the other vertices *j* have $reachable_1(j)=false$). Then, we perform a *BFS* from the sink(s), considering the reverse direction of the edges. As before, we only consider those edges *(u->v)* with *f(u->v)<cap(u->v)*. We mark the reachable vertices *i* by $reachable_2(i)=true$ (the other vertices *j* have $reachable_2(j)=false$). A directed edge *(u->v)* is upward critical if *f(u->v)=cap(u->v)* and $reachable_1(u)=true$ and $reachable_1(v)=false$ and $reachable_2(v)=true$ and $reachable_2(u)=false$. If the multigraph is undirected then: *(1)* we compute a maximum flow ; *(2)* we consider only the edges *(u,v)* with *f(u,v)<cap(u,v)* when performing the two breadth-first searches ; *(3)* for every edge *(u,v)*, we consider its two possible orientations, *(u->v)* and *(v->u)*, and use the criterion described above; *(u,v)* is upward critical if it is classified as such for one of the two orientations.

## 4. Conclusions and Related Work

In this paper we addressed several network design and analysis problems, for which we presented novel algorithmic solutions. Similar or related problems were considered in [2, 4]. Some of our solutions are based on algorithmic techniques presented in [1, 3].